\documentclass[]{aa}  

\usepackage{supertabular,booktabs}
\usepackage{graphicx}
\usepackage{lscape}
\usepackage{natbib}
\usepackage{bigstrut} 
\usepackage{amsmath}
\usepackage{epstopdf}
\usepackage[varg]{txfonts}
\begin{document} 

\title{Search for systemic mass loss in Algols with bow shocks}

\author{A.~Mayer
                \inst{1}
                \and R.~Deschamps
                \inst{2,3} 
                \and A.~Jorissen\inst{2}}

\institute{University of Vienna, Department of Astrophysics, Sternwartestra\ss e 77, 1180 Wien, Austria \\
\email{a.mayer@univie.ac.at}
\and
Institut d'Astronomie et d'Astrophysique, Universit\'e Libre de Bruxelles CP 226, Av. F. Roosevelt 50,  B-1050 Brussels, Belgium
\and
European Southern Observatory, Alonso de Cordova 3107, Casilla 19001, Santiago, Chile}

\date{Received 28 May 2015 ; accepted 24 December 2015}

\abstract
  % context heading (optional)
{}
  % aims heading (mandatory)
{Various studies indicate that interacting binary stars of Algol type evolve non-conservatively. However, direct detections of systemic mass loss in Algols have been scarce so far. We  study the systemic mass loss in Algols by looking for the presence of infrared excesses originating from the thermal emission of dust grains, which is linked to the presence of a stellar wind.}
  % methods heading (mandatory)
{In contrast to previous studies, we make use of the fact that stellar and interstellar material is piled up at the edge of the astrosphere where the stellar wind interacts with the interstellar medium. We analyse WISE W3 $12\,\mu$m and WISE W4 $22\,\mu$m data of Algol-type binary Be and B[e] stars and the properties of their bow shocks. From the stand-off distance of the bow shock we are able to determine the mass-loss rate of the binary system.}
  % results heading (mandatory)
{Although the velocities of the stars with respect to the interstellar medium are quite low, we find bow shocks  present in two systems, namely \object{$\pi$~Aqr}, and \object{$\varphi$~Per}; a third system, \object{CX~Dra}, shows a more irregular circumstellar environment morphology which might somehow be related to systemic mass loss. The properties of the two bow shocks point to mass-loss rates and wind velocities typical of single B stars, which do not support an enhanced systemic mass loss.}
{}


\maketitle

\section{Introduction}

The group of Algols host stars with many different observed properties, like W~Ser stars, $\beta$~Lyr{\ae} stars, binary B[e] and Be stars, and symbiotic Algols, which all have in common the paradox   that the donor star is more evolved but less massive than the accretor. This is achieved by mass transfer when at a certain point the mass ratio  reverses. Non-conservative evolution in Algol-type binary systems has been known for 60 years \citep{1955ApJ...121...71C}. For example  \citet{Chaubey1979}, \citet{Sarna1993}, and \citet{vanRensbergen2011} noted that Algol models must lose a significant fraction of their mass to reproduce observed properties. One of the most efficient scenarios that removes mass from the system is via a hotspot on the surface of the gainer\footnote{for an extensive explanation of the hotspot mechanism, see \citet{vanRensbergen2011} and \citet{Deschamps2013,Deschamps2015}.}. However, no direct detection of systemic mass loss during the mass transfer process in close binaries has been reported for Algols so far. 

In this work, we focus on Be and B[e] stars for which binarity has been confirmed and the properties of the systems are well constrained. A Be~star is a non-supergiant B~star whose spectrum has, or had at some time, one or more Balmer lines in emission, and might also show infrared (IR) excess \citep{Collins1987}. The origin of these spectral features in these binary Be stars is probably linked to the mass-transfer event \citep{2003PASP..115.1153P}. The IR excess is most likely caused by hot circumstellar dust \citep{1998A&A...340..117L}, which is present in the case of an evolved binary system associated with mass transfer events \citep[e.g.][]{2012A&A...542A..50D}. B[e] stars also have   strong forbidden (Fe) emission lines. 

We take advantage of the fact that some of these stars are not at rest, but move with a certain speed with respect to their surrounding medium. Assuming that mass is expelled supersonically from the binary system, it is decelerated by the oncoming interstellar medium (ISM) and forms a bow shock \citep{Baranov1971,Weaver1977}. Bow shocks have been observed at all kinds of wavelengths around many stellar types, covering runaway O stars to AGB stars \citep[e.g.][]{vanBuren1988,Cox2012}. In the mid-IR, these shocks are visible through thermal dust emission, when the shock front heats up dust grains at the interface between the stellar wind and the ISM \citep{Ueta2006}. A bow shock detection around an Algol is therefore  direct evidence of stellar material around the binary system. In this case, the distance of the system to the apex of the bow shock can be used to derive the systemic mass-loss rate if the wind velocity, stellar velocity, and ISM density are known \citep{Baranov1971}. 

This research note is a complementary study to the work done by \citet[][hereafter D15]{Deschamps2015}, but with emphasis on the observational aspects of the systemic mass loss. Based on radiative transfer calculations, \citet[][their Fig.~13]{Deschamps2015} predicted the IR colour excesses expected in the case of systemic mass loss. In addition to the WISE detections of extended material around CZ~Vel and SX~Aur presented in D15, we discuss the properties of the circumstellar emission of three other objects, namely CX~Dra, $\pi$~Aqr, and $\varphi$~Per.

\section{WISE observations}
\label{sec:wise}

\begin{figure}[t]
  \centering
  \includegraphics[width=9cm]{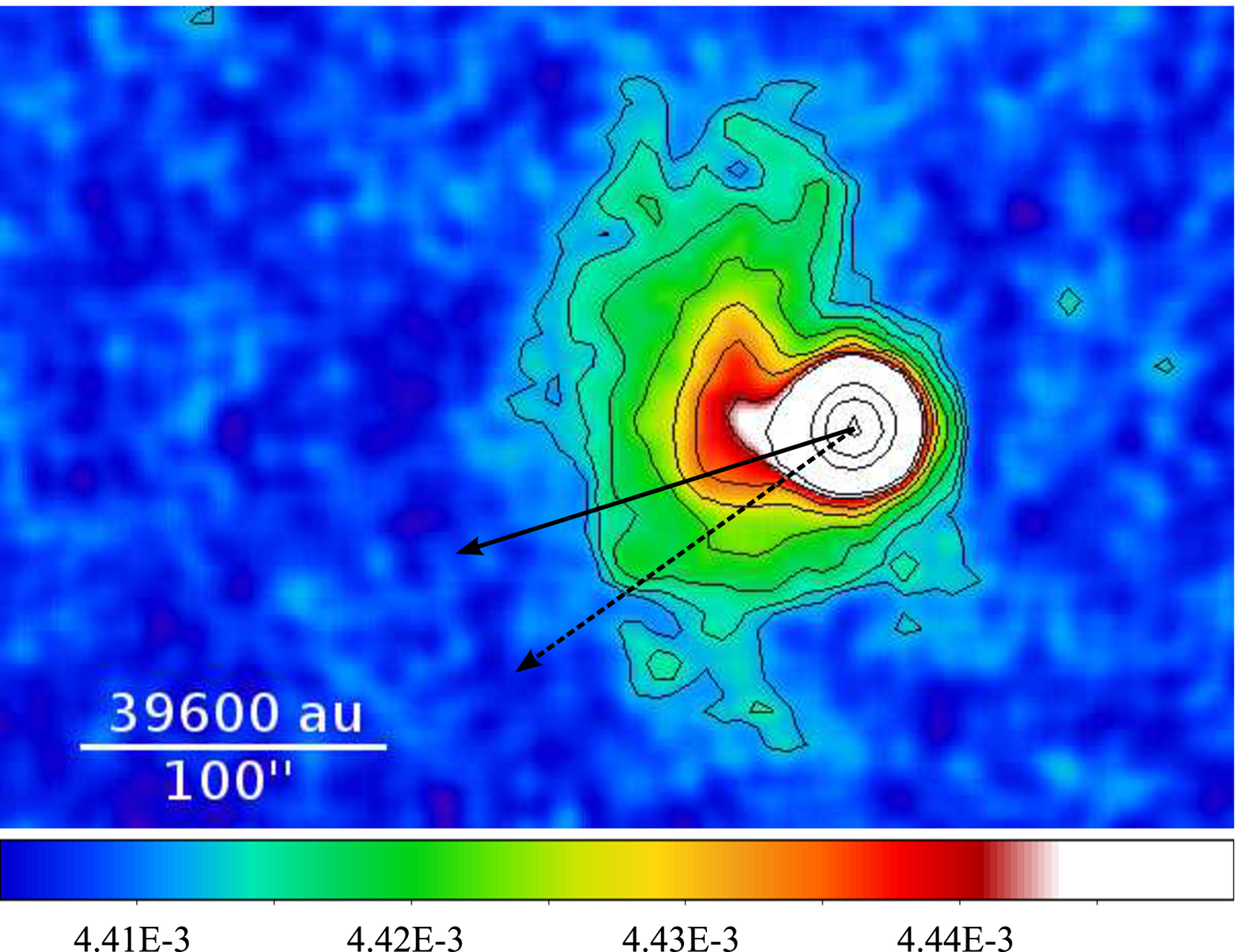}
  \includegraphics[width=9cm]{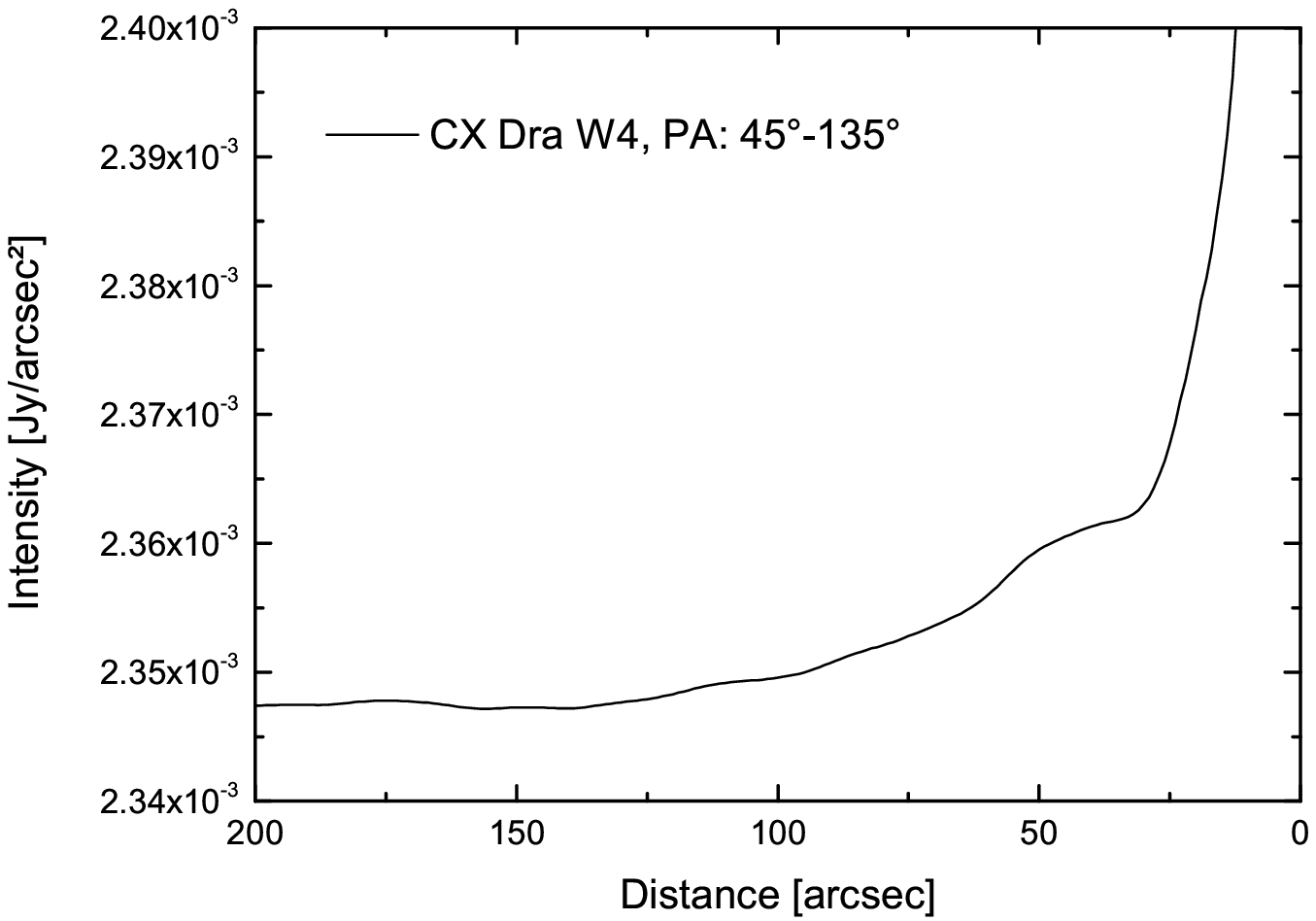}
  \caption{\textit{Upper panel}: WISE W4 image of CX~Dra at $22\,\mu$m. The continuous black arrow gives the uncorrected proper motion from the reprocessed Hipparcos catalogue \citep{vanLeeuwen2007}, while the dashed arrow points to the direction of the space motion corrected from the solar motion \citep{Coskunoglu2011}. The values of the motion are given in Tab.~\ref{tab:pa}. The values of the colour bar are given in Jy\,pix$^{-1}$. \textit{Lower panel}: Integrated intensity cut through a wedge covering position angles (P.A.):~45$\degr$--135$\degr$.}
  \label{fig:cx_dra}
\end{figure}

\begin{figure}[t]
  \centering
  \includegraphics[width=9cm]{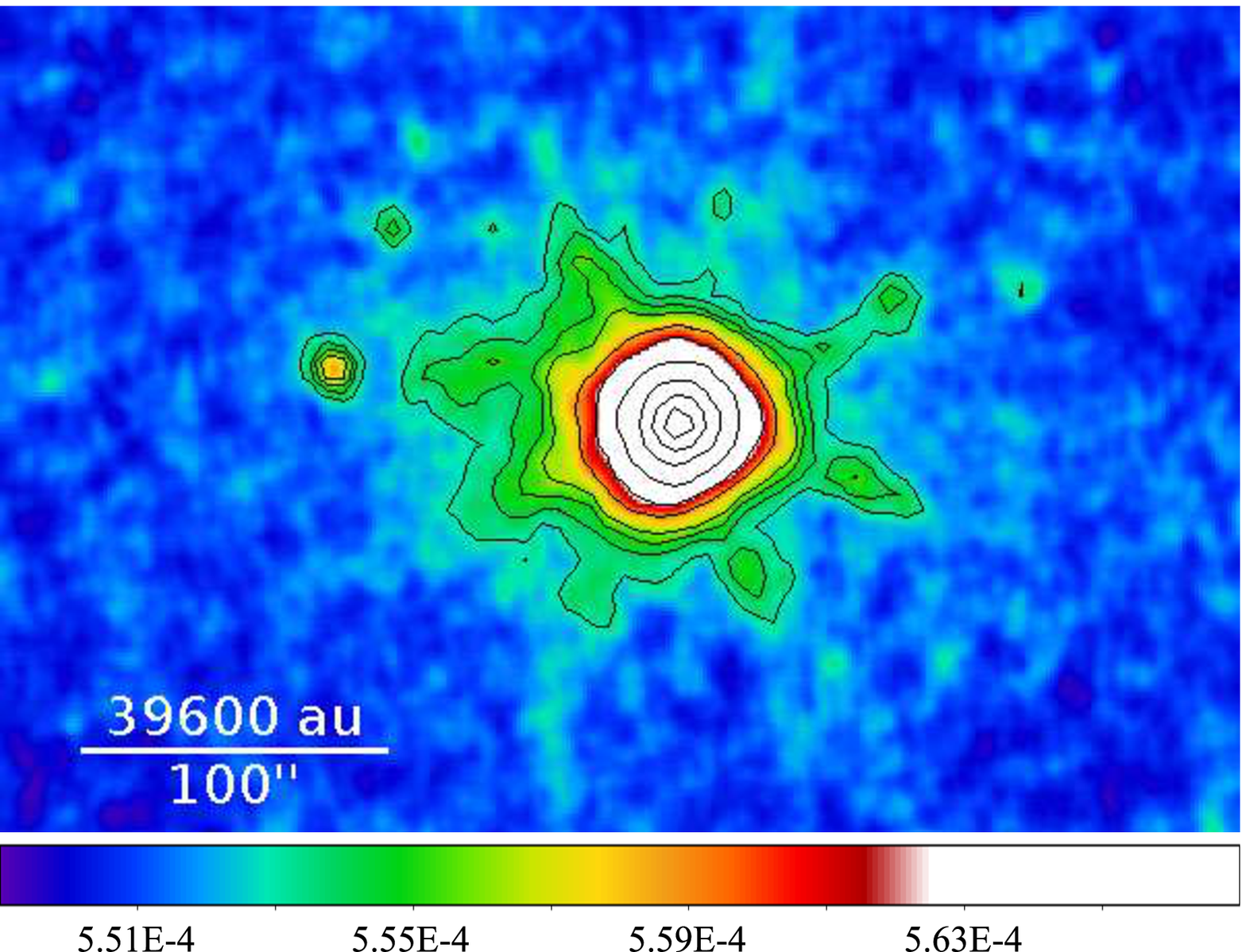}
  \includegraphics[width=9cm]{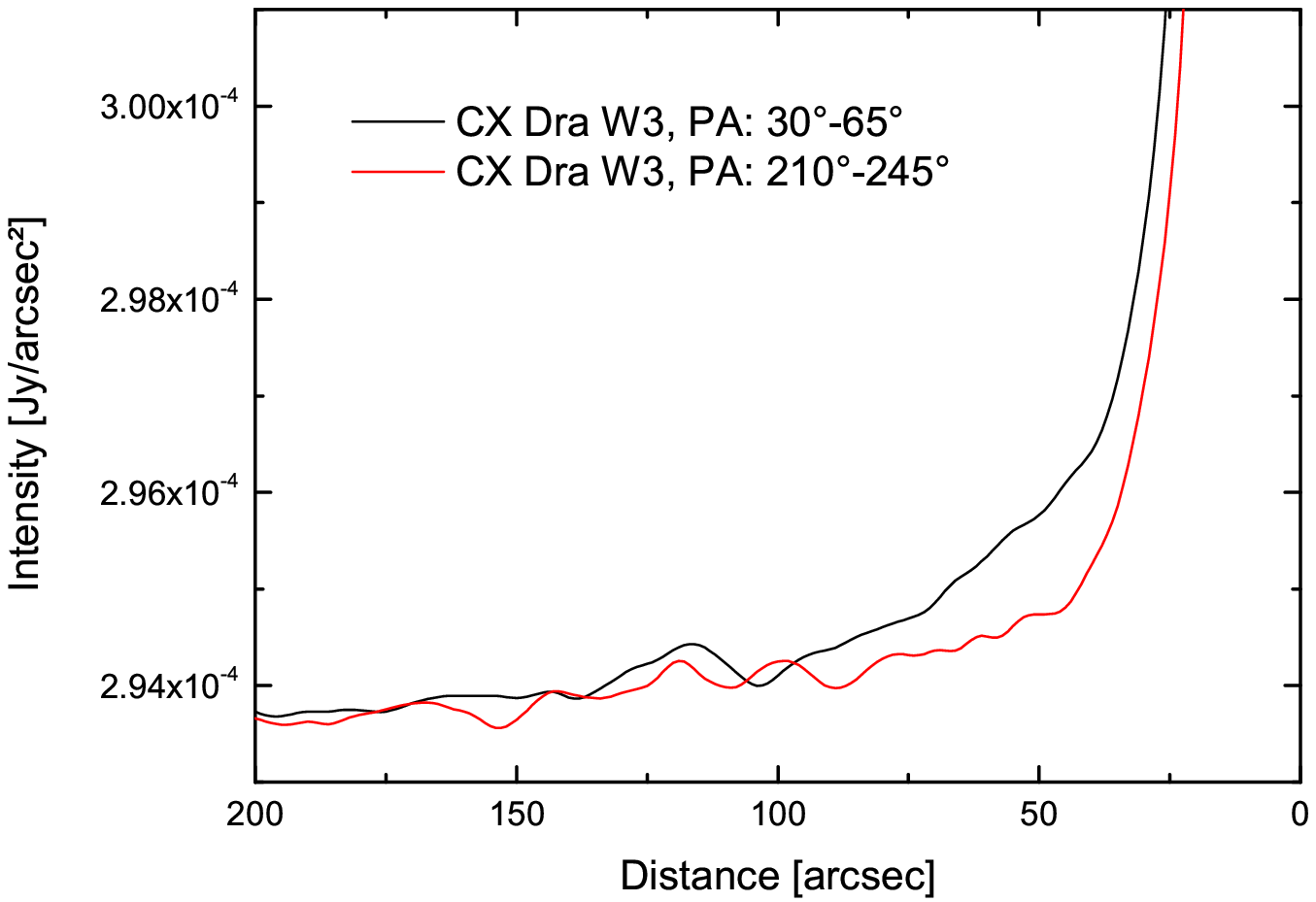}
  \caption{\textit{Upper panel}: WISE W3 images of CX~Dra at $12\,\mu$m. \textit{Lower panel}: Integrated intensity cuts through a wedge covering P.A.:~30$\degr$--65$\degr$ (black line) where emission is visible and P.A.:~210$\degr$--245$\degr$ (red line) without extended emission. We had to choose smaller wedges to avoid the flux being dominated by the diffraction spikes of the PSF.}
  \label{fig:cx_dra_W3}
\end{figure}

\begin{figure}[t]
  \centering
  \includegraphics[width=9cm]{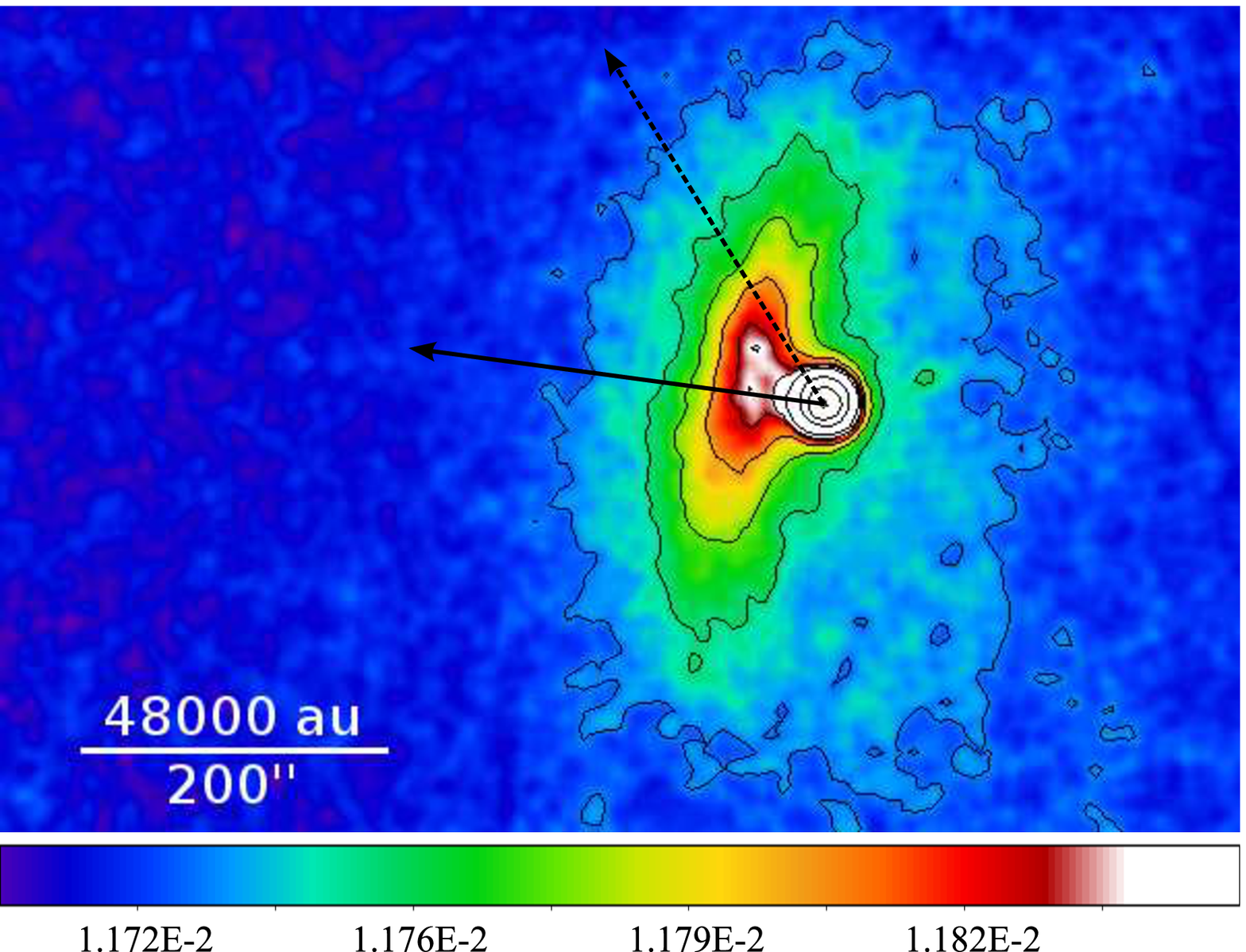}
  \includegraphics[width=9cm]{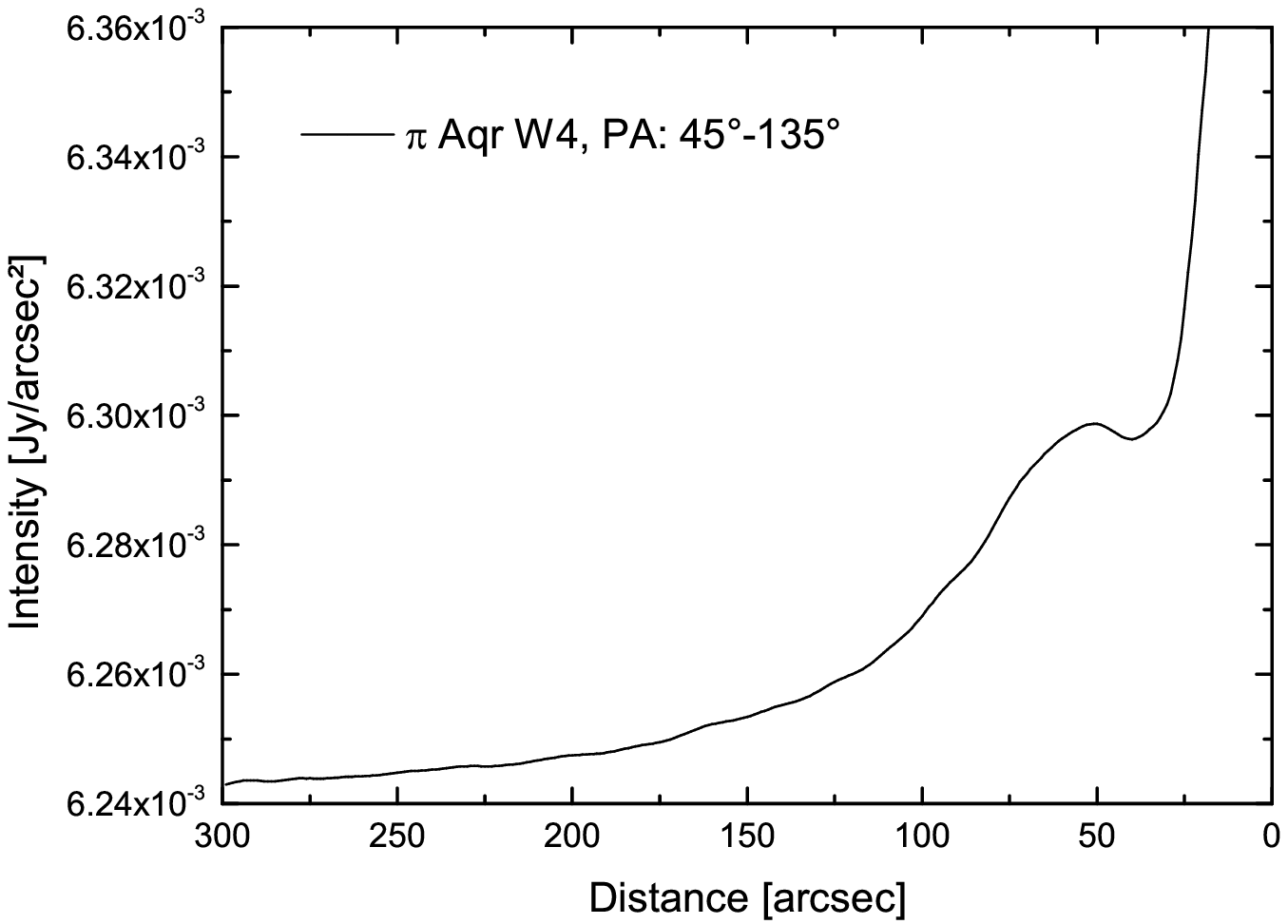}
  \caption{Same as Fig.~\ref{fig:cx_dra} for $\pi$~Aqr. The integrated intensity cut in the lower panel covers PA:~45$\degr$--135$\degr$.}
  \label{fig:pi_aqr}
\end{figure}

\begin{figure}[t]
  \centering
  \includegraphics[width=9cm]{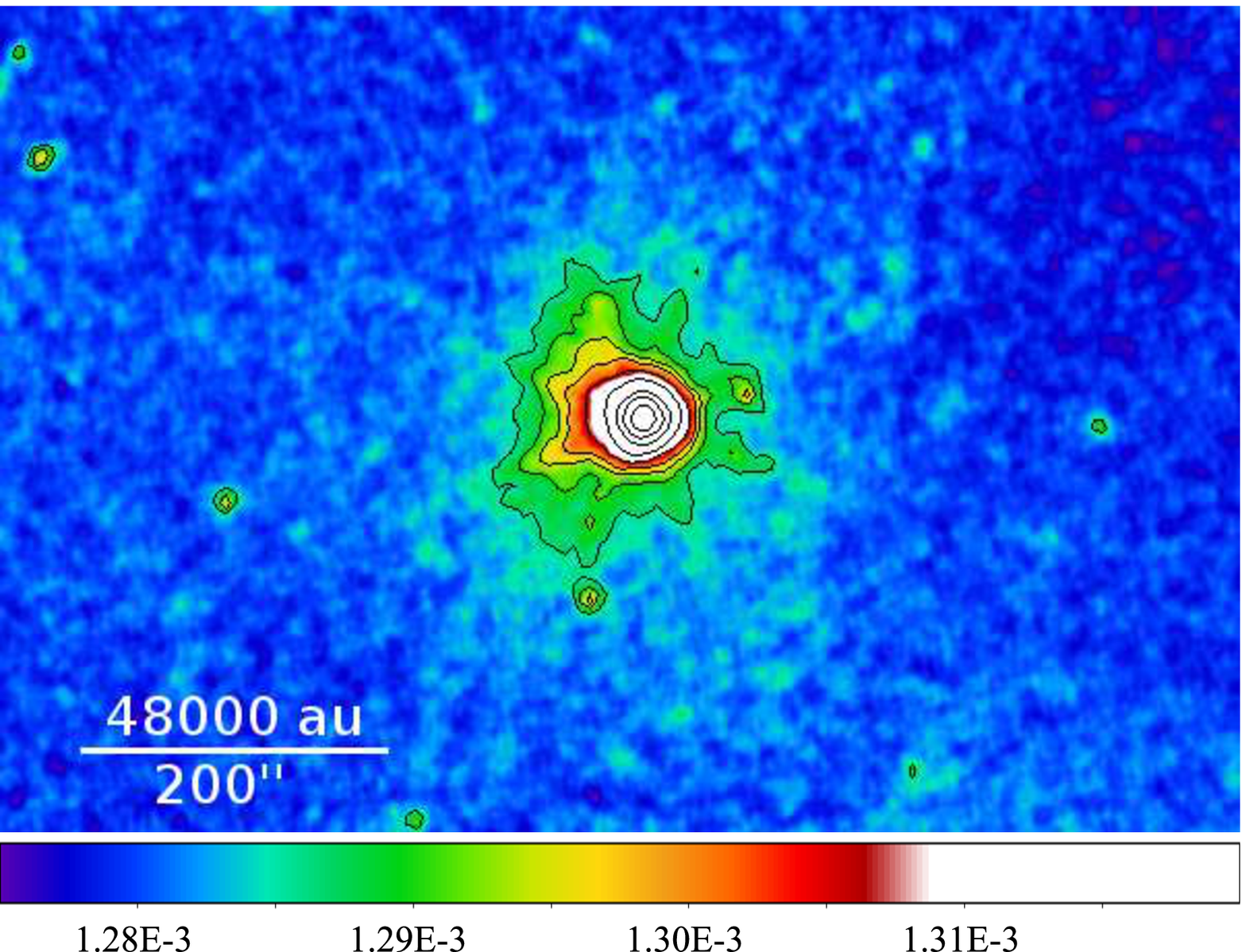}
  \includegraphics[width=9cm]{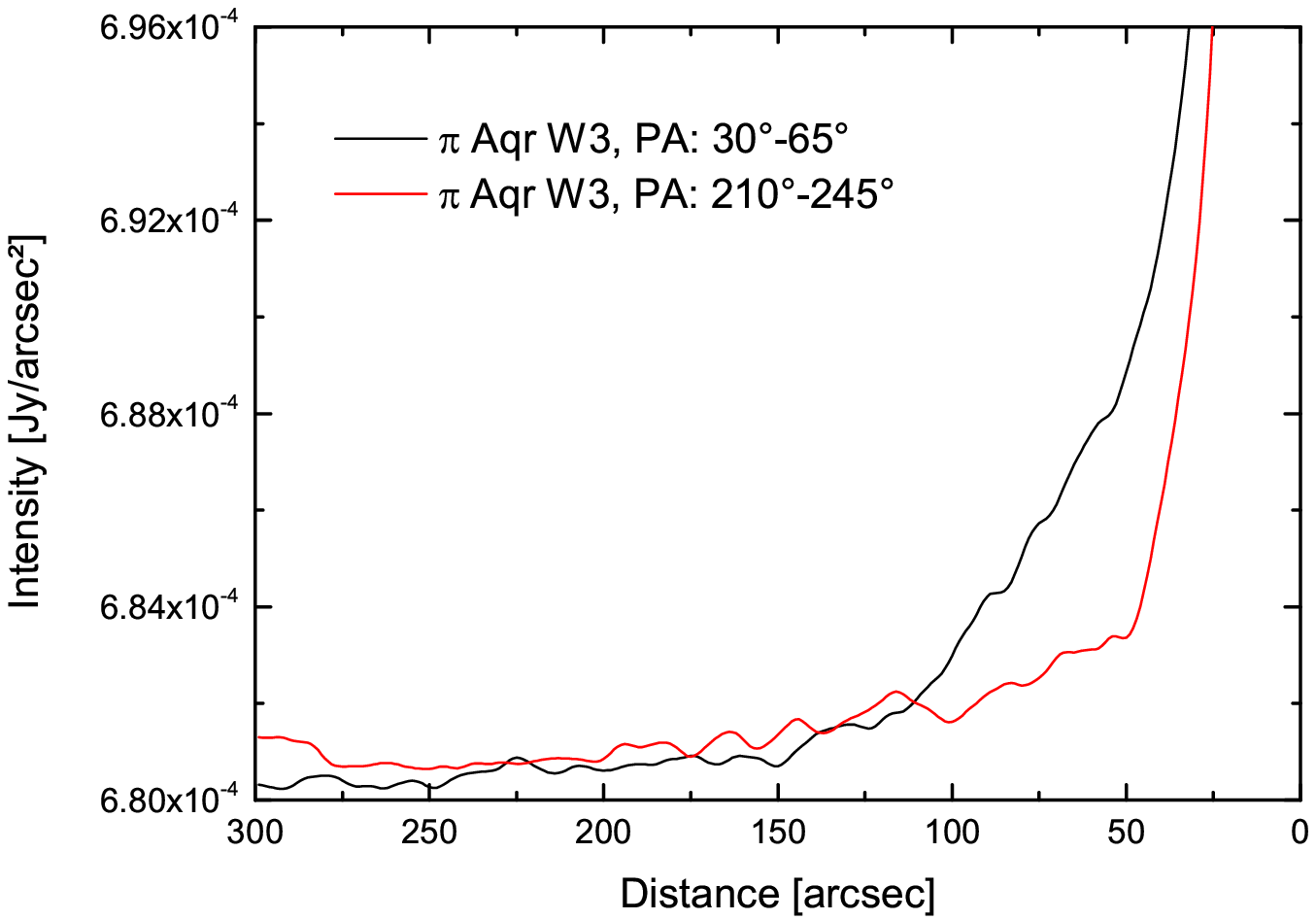}
  \caption{Same as Fig.~\ref{fig:cx_dra_W3} for $\pi$~Aqr at 12\,$\mu$m. The integrated intensity cuts in the lower panel cover P.A.:~30$\degr$--65$\degr$ (black line) where emission is visible and P.A.:~210$\degr$--245$\degr$ (red line) without extended emission.}
  \label{fig:pi_aqr_W3}
\end{figure}

\begin{figure}[t]
  \centering
  \includegraphics[width=9cm]{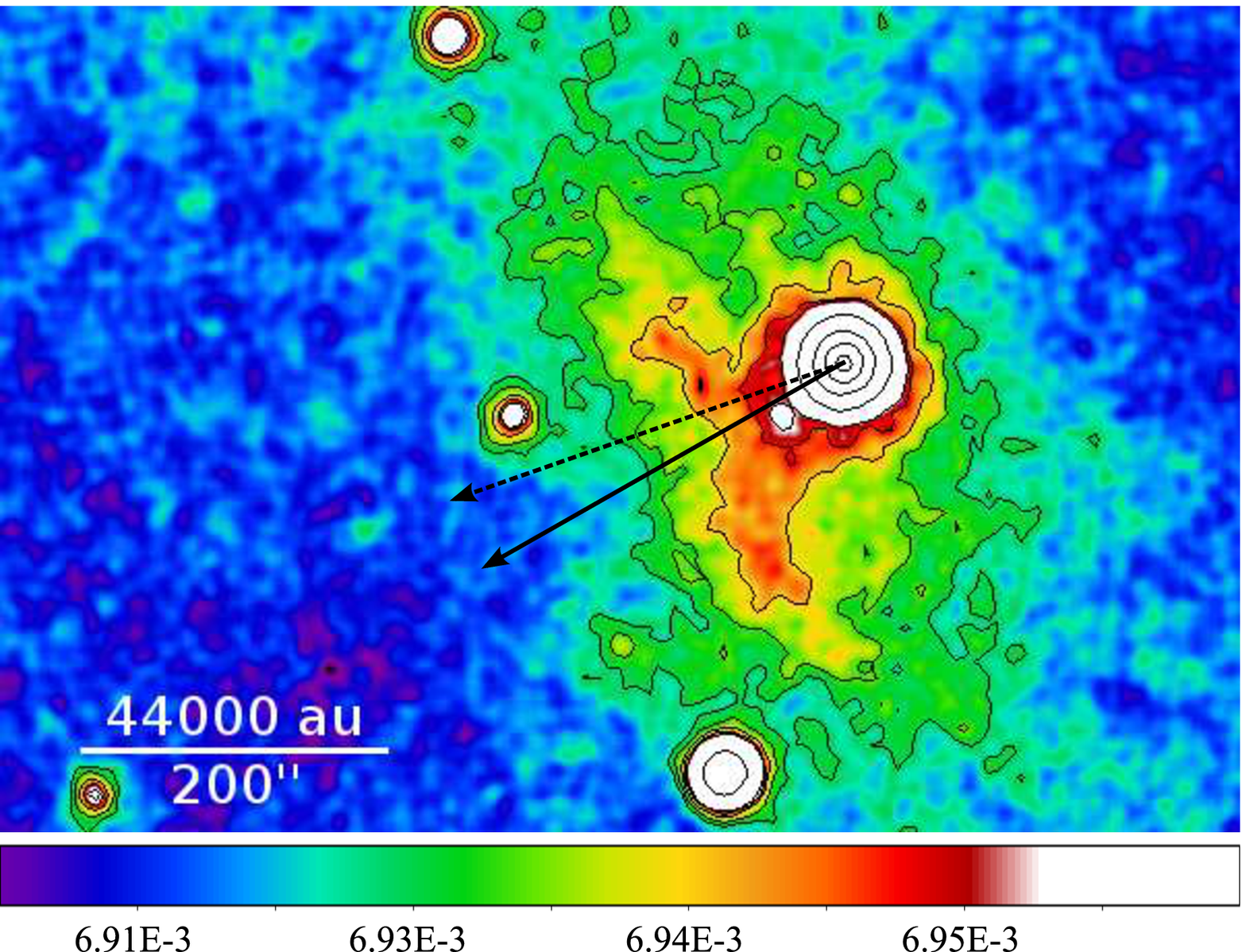}
  \includegraphics[width=9cm]{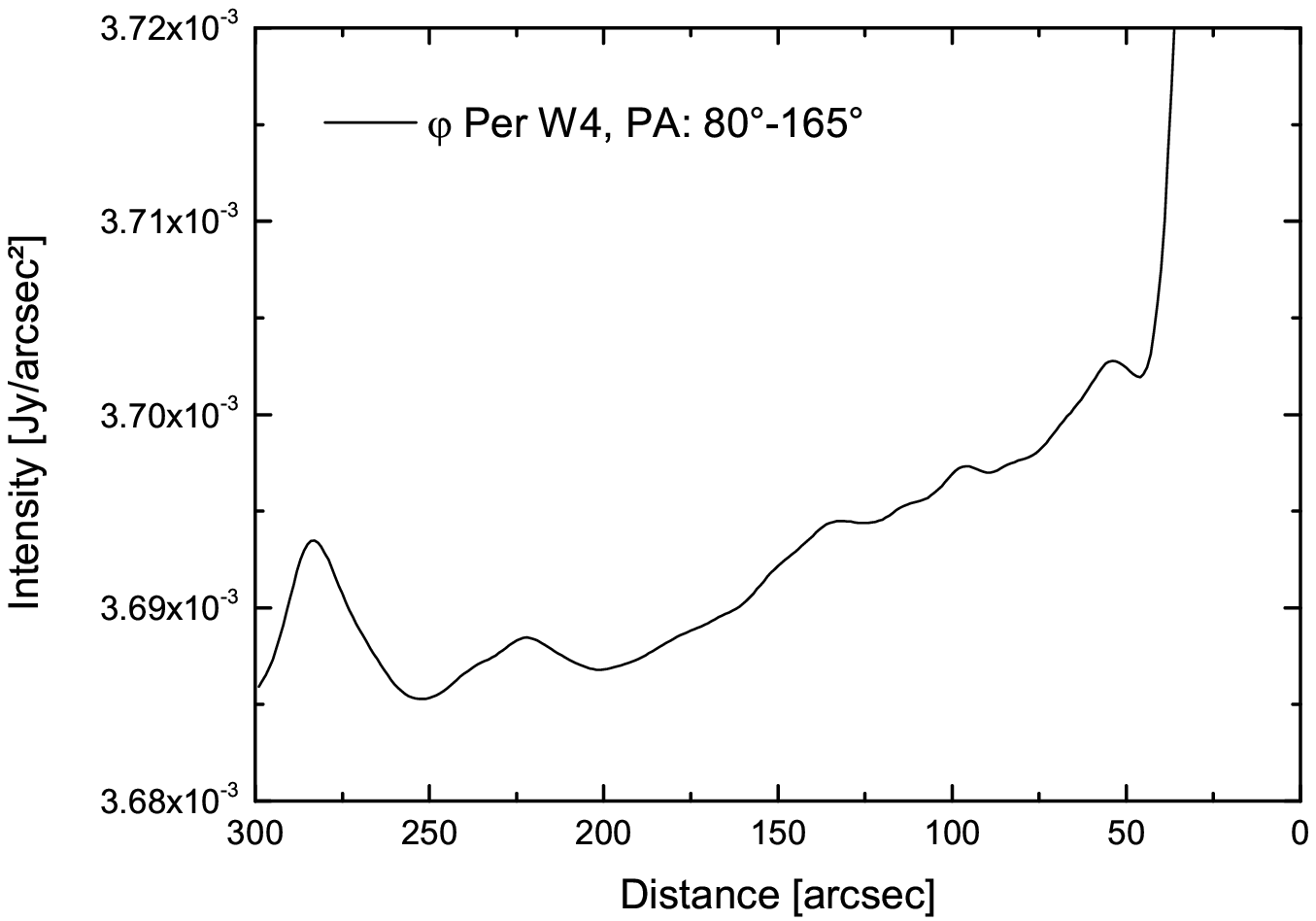}
  \caption{Same as Fig.~\ref{fig:cx_dra} for $\varphi$~Per. The integrated intensity cut in the lower panel covers P.A.:~80$\degr$--165$\degr$. The peaks at distances of about $223\arcsec$ and $283\arcsec$ in the lower panel correspond to the stars visible in the image at position angles of $99\degr$ and $164\degr$, respectively.}
  \label{fig:phi_per}
\end{figure}

In D15, we performed a systematic search for extended IR emission around Algols (collected from the catalogues of Brancewicz \& Dworak 1980 and  Budding et al. 2004) \nocite{Brancewicz1980,Budding2004} and Algol-related Be and B[e] systems \citep{Harmanec2001}, using archive data from the Wide-field Infrared Survey Explorer (WISE)\footnote{The IRSA:WISE archive can be found at \url{http://irsa.ipac.caltech.edu/applications/wise/}}. WISE is an all-sky survey, which mapped the sky in four bands at 3.4, 4.6, 12, and 22\,$\mu$m with angular resolutions of 6$\farcs$1, 6$\farcs$4, 6$\farcs$5, and 12$\farcs$0, respectively \citep{Wright2010}. Based on the list of 70 objects (Algols and Algol-like Be stars with a WISE-source counterpart) provided by D15\footnote{The list of objects can be found in Appendix~\ref{objectlist}}, we found that three systems, CX~Dra, $\pi$~Aqr, and $\varphi$~Per,  have unambiguous circumstellar emission and  therefore deserve a specific analysis (in addition to  CZ~Vel and SX~Aur, already discussed in D15). 

All three objects are in the list of binary Be stars compiled by \citet{Harmanec2001}. They exhibit peculiarities that flag them as Algol candidates, or at least as systems with on-going mass transfer. With its sdO companion, $\varphi$~Per has obviously undergone a severe mass transfer, the primary and more luminous B2[e] component being the most massive but the least evolved. In CX~Dra, mass transfer in the binary has been inferred from the variation of the polarisation with orbital phase, and from the fact that the less massive, more evolved F5III companion fills its Roche lobe \citep{Berdyugin2002}. In $\pi$~Aqr photometric, spectroscopic (broad and complex H$\alpha$ line profile), and polarimetric variations observed during the second half of the 20th century are tentatively attributed to variable mass transfer between the binary components \citep{Bjorkman2002,Hanuschik1996}.

The extended emission around the stars was detected in Band~3 (W3) at 12\,$\mu$m (CX~Dra, $\pi$~Aqr) and Band~4 (W4) at 22\,$\mu$m (CX~Dra, $\pi$~Aqr, $\varphi$~Per). For the two objects with circumstellar emission (CSE) detected in both bands, WISE W4 offers greater details, most likely because the thermal emission of the shock-heated dust grains peaks at longer wavelengths \citep{Draine1981}. In the following, the CSM morphology of the three objects is described. Figures~\ref{fig:cx_dra}--\ref{fig:phi_per} depict the WISE images of CX~Dra, $\pi$~Aqr, and  $\varphi$~Per, while Tab.~\ref{tab:pa} provides their stellar properties.

\subsection{CX~Dra}

CX~Dra (HIP~92133) is a 6.696\,d period Algol B2.5Ve+F5III system at a distance of 396\,pc \citep{vanLeeuwen2007};  one of the components rotates rapidly. Although it is  not  eclipsing, \cite{Berdyugin2002} estimate the mass of the two components to be $3.9\,M_{\odot}$ and $0.9\,M_{\odot}$ at $i=70\degr$. The authors, however, correctly note that these masses are too small to match the spectral types of the two stars.  

The circumstellar environment of the star is shown in Figs.~\ref{fig:cx_dra} and \ref{fig:cx_dra_W3}. The emission in the WISE W4 $22\,\mu$m band is concentrated to the east of the star and is traceable to a distance of about $120\arcsec$ (47\,500\,au at 396\,pc). The morphology of the circumstellar material is somewhat puzzling because several aspects are not in favour of an ISM interaction. First, the direction of the proper motion is S-E, but the shape of the emission is not symmetric and is more concentrated N-E of the star. Second, the emission is not detached from the star and the flux seems to decrease with distance, which is not  expected for a bow shock where the brightest region is at the position of the shock front. 

Furthermore, the circumstellar material of CX~Dra on the WISE image describes an arc emerging east of the star and curved towards the north. Similar arcs are found to be part of an Archimedean spiral which is caused by a semi-detached companion interacting with the wind of a primary \citep[e.g.][]{Mastrodemos1999,Mayer2011,Maercker2012}. The spacing of the spiral arms is thereby defined by the wind velocity of the mass losing star and the orbital period of the companion. However, assuming a wind velocity of 1000\,km\,s$^{-1}$ for the B2.5Ve star and an orbital period of 6.696\,d for the F5III companion, the resulting spiral spacing is 3.87\,au, which is several orders of magnitude smaller than what is seen on the WISE image. For comparison, the pixel size of the image is $1\farcs375$, which is 545\,au at 396\,pc. This implies that a spiral formed by the B2.5Ve+F5III system would show 140 windings per WISE W4 pixel. The observed arc is therefore not related to this shaping mechanism.

In the colour-colour diagram shown in Fig.~13 of D15 that depicts the WISE W4/W1 against 2MASS J/K$_{\rm s}$ flux ratios, CX~Dra is located only slightly above the black-body  curve (F$_{\rm J}$/F$_{\rm K_s}$=1.678, F$_{\rm W4}$/F$_{\rm W1}$=0.047). Many other objects fall into this region of the diagram and no peculiarity can be drawn from it. Still, there is no doubt that extended emission is present in the WISE W4 image. 

In the shorter WISE W3 band at 12\,$\mu$m (see Fig.~\ref{fig:cx_dra_W3}), CX~Dra also shows extended emission east of the star but in much less detail than in the W4 image. The detection of emission in W3 and W4, however, allows us to estimate the temperature of the dust emission around CX~Dra. We performed aperture photometry on a circle of radius $15\arcsec$ in both bands. The region we chose is centred at a distance of $56\arcsec$ from the star at PA~$=48\degr$ and falls between the diffraction spikes of the PSF which is dominating the W3 image. The resulting fluxes are $F_{\nu,12}= 0.207$\,Jy and $F_{\nu,22}= 1.650$\,Jy at 12\,$\mu$m and 22\,$\mu$m, respectively. Adopting the corresponding absorption coefficients of astronomical silicates $Q_{\rm abs,12} = 5.60 \times 10^{-2}$ and $Q_{\rm abs,22} = 3.39 \times 10^{-2}$ \citep{Draine1985}, the 12\,$\mu$m and 22\,$\mu$m fluxes correspond to a temperature of 124\,K \citep[for details see][]{Jorissen2011}.

Since no other archival observations are available for CX~Dra, we cannot conclude on the shaping mechanism of its circumstellar material. We note, however,  that the star might be a possible candidate for showing systemic mass loss in its circumstellar environment, but further observations are needed.

\subsection{$\pi$~Aqr}

Pi Aquarii (HIP~110672) is a 84.1\,d period binary located 240\,pc  from the sun \citep{vanLeeuwen2007}. The system comprises a rapidly rotating B1Ve star at the origin of the Be phenomenon and an A-F type companion. \cite{Bjorkman2002} estimate the mass of the components to be $M_{1} \sin^{3} i = 12.4\,M_{\odot}$ and $M_{2} \sin^{3} i = 2.0\,M_{\odot}$ with an orbital inclination $i = (50-75)^{\circ}$. The stellar wind is one of the fastest among the Be stars with a terminal velocity of 1450\,km\,s$^{-1}$;  the mass-loss rate is one of the highest with $\dot{M} = 2.61\times10^{-9}\,M_{\odot}$\,yr$^{-1}$ estimated from the Si\,\textsc{iv} profile \citep{Snow1981}. We note that the Si\,\textsc{iv} lines  likely  form inside the Roche lobe of the gainer star and might therefore not trace the systemic mass-loss rate. 

The WISE W4 image of $\pi$~Aqr is depicted in Fig.~\ref{fig:pi_aqr}. The emission shows a morphology that is typical for a wind-ISM interaction with a bow shock in the direction of the space motion. The bow shock cone is quite symmetric on the northern and southern half, extending to about $220\arcsec$ (52\,800\,au at 240\,pc) in those directions. In the direction of motion, the material can be traced to about $150\arcsec$ (36\,000\,au) from the binary system. The emission peak, however, is closer to the system at about $52\arcsec$ (12\,480\,au).

In WISE W3 at $12\,\mu$m (see Fig.~\ref{fig:pi_aqr_W3}), the CSE is concentrated to the east of the star at the same position where the bow shock in W4 is visible, but not as extended in the north-south direction. The lower panel of Fig.~\ref{fig:pi_aqr_W3} shows cuts through regions with and without extended emission.

In the same manner as for CX~Dra, we also performed aperture photometry ($r=15\arcsec$)  for $\pi$~Aqr in both bands at a region centred at a distance of $61\arcsec$ from the star at PA~$=46\degr$. The resulting fluxes are $F_{\nu,12}= 0.481$\,Jy and $F_{\nu,22}= 4.411$\,Jy at 12\,$\mu$m and 22\,$\mu$m, respectively. Adopting the same absorption coefficients of astronomical silicates as for CX~Dra, the 12\,$\mu$m and 22\,$\mu$m fluxes correspond to a temperature of 120\,K.

\subsection{$\varphi$~Per}

Phi Persei (HIP~8068) is a long period Algol B2[e]+sdO system ($P_{\mathrm{orb}}=127$\,d) at a distance of 220\,pc \citep{vanLeeuwen2007}. The system is likely at the end of its mass-transfer phase \citep{1998ApJ...493..440G} and the material transferred from the donor star has largely spun up the gainer star (primary) to the rotation rate now observed. Based on double-line spectroscopic orbital elements, the masses of the components have been estimated to be $9.3\pm0.3\,M_{\odot}$ for the B[e] primary and $1.14\pm0.04\,M_{\odot}$ for the sdO secondary (donor star). A hotspot region detected on the edge of the disc produces strong Fe\,\textsc{iv} lines. The envelope of the companion has mostly been stripped off by the Roche-lobe overflow (RLOF) event and the secondary, now a hot sdO star, is only visible in the UV. 

The WISE W4 $22\,\mu$m emission of $\varphi$~Per is shown in Fig.~\ref{fig:phi_per}. The CSM is elliptically shaped with the major axis approximately in the N-S direction. The extent of the emission to the south is about $290\arcsec$ (63\,800\,au at 220\,pc), while the frame is cut off in the north $250\arcsec$ from the star. East of the star at $\approx 100\arcsec$ (22\,000\,au), a brightened bar is visible with the same N-S orientation as the whole elliptical emission and a length of about $240\arcsec$ (52\,800\,au). The bar is bent towards the star at the same position angle as the direction of the space motion, which indicates that this is the interface where the ISM interacts with the stellar material. These bendings are visible in hydrodynamic simulations of bow shocks where the shocked stellar and ambient material cool efficiently \citep[see Fig.~15 in][]{Comeron1998}. A beautiful  example of  a bent bow shock is found around the AGB star X~Her \citep{Jorissen2011}. In contrast to the other two objects, $\varphi$~Per does not show extended emission in WISE W3.

\section{Bow shock properties and systemic mass loss}
\label{bow_shocks}

\begin{table}[t!]
  \centering
  \caption{Stellar properties. The space velocities were calculated following \citet{Johnson1987} using the Hipparcos proper motion and parallax \citep{vanLeeuwen2007} and radial velocities from the catalogue by \citet{deBruijne2012}. The  solar motion adopted to convert heliocentric motion into LSR motion is $(U,V,W) = (8.50\pm0.29$, $13.38\pm0.43$, $6.49\pm0.26$) km\,s$^{-1}$ \citep{Coskunoglu2011}.}
  \label{tab:pa}
  \begin{tabular}{l|rrrrrrrrrr}
    \hline
    \hline
 & CX~Dra &  $\pi$~Aqr & $\varphi$~Per \\
 \hline
 Spec. type & B2.5Ve+F5III & B1Ve+[A-F] &  B2[e]+sdO \\
 $M_1$ [$M_\sun$] & 3.9 & $14.0\pm1.0$ & $9.3\pm0.3$ \\
 $M_2$ [$M_\sun$] & 0.9 & $2.3$ & $1.1\pm0.1$ \\
 $P_{\rm orb}$ [d] & 6.696 & 84.1 & 127 \\
 D [pc] & $396\pm35$ & $240\pm15$ & $220\pm9$ \\
 z [pc] & 149 & 169 & 43 \\
$n_{\rm H}$ [cm$^{-3}$] & 0.45 & 0.37 & 1.30 \\
 RV [km/s] & $-2.1\pm2.3$ & $-4.9\pm0.1$ & $-4.0\pm2.1$ \\
 \hline
 $v_*$ [km/s] & $21.1\pm2.0$ & $21.0\pm1.1$ & $29.8\pm1.6$ \\
 P.A. [$\degr$] & $107.6\pm1.2$ & $82.3\pm0.6$ & $119.7\pm0.2$ \\
 $i$ [$\degr$] & $-5.7\pm63.1$ & $-13.5\pm1.2$ & $-7.7\pm30.3$ \\
 \hline
 $v_{\rm *,LSR}$ [km/s] & $25.5\pm2.1$ & $13.5\pm1.0$ & $13.1\pm1.7$ \\
 P.A.$_{\rm LSR}$ [$\degr$] & $125.7\pm1.0$ & $31.5\pm0.5$ & $109.4\pm0.6$ \\
 $i_{\rm LSR}$ [$\degr$] & $32.4\pm11.4$ & $6.6\pm3.7$ & $-4.4\pm90$ \\
 \hline
 Ref. & 1 & 2,3,4 & 5 \\
 \hline
   \end{tabular}
     \tablebib{(1)~\citet{Berdyugin2002}; (2)~\citet{Zharikov2013};
    (3)~\citet{Linnell1988}; (4)~\citet{Bjorkman2002};
    (5)~\citet{Hummel2001}}
\end{table}

For a star that moves with respect to its surrounding medium, the stellar motion adds an asymmetry to the wind velocity profile, since different parts of the wind face the ISM with different relative velocities. If the motion is supersonic, a bow shock arises at the interface where the ram pressure of the ISM and the stellar wind balance. The stand-off distance, i.e. the distance of the star to the apex of the shock front, is given by
\begin{equation}
\label{R0_eq}
R_0=\sqrt{\frac{\dot{M} v_w}{4 \pi \rho_0 v_*^2}},
\end{equation}
where $v_w$ is the terminal wind velocity, $v_*$ the stellar velocity with respect to the ISM, $\dot{M}$ the mass-loss rate, and $\rho_0$ the density of the surrounding medium \citep{Baranov1971}. The density can be expressed in  number density of hydrogen atoms ($m_{\rm H} = 1.6727\times 10^{-27}$\,kg), which follows roughly
\begin{equation}
\label{density}
n_{\rm H} = 2.0 \; e^{-\frac{|z|}{100\,{\rm pc}}},
\end{equation} 
where $z$ is the galactic height \citep{Mihalas1981} and $n_{\rm H}$ is given in atoms per cm$^3$. \citet{Wilkin1996} demonstrated that the shape of the bow shock only depends on the stand-off distance, while \citet{Cox2012} showed that this assumption remains valid for viewing angles up to $70\degr$. Above this value, the bow shock cone becomes broader. Therefore, we were able to use Eq.~\ref{R0_eq} to estimate the mass-loss rate from the binary system by measuring the stand-off distance. Generally, the ISM density and stellar velocity can be determined following Eq.~\ref{density} and \citet{Johnson1987}, respectively. While the error of the space motion is negligible, the ISM density value is only an estimate since the star could move through a dense cloud, which is not considered by Eq.~\ref{density}. The respective values of these quantities for the three objects are given in Tab.~\ref{tab:pa}. To obtain the space motion ($v_{\rm *,LSR}$) with respect to the local standard of rest, we corrected the heliocentric motions from the solar motion vector $(U,V,W)_\sun = (8.50\pm0.29, 13.38\pm0.43, 6.49\pm0.26)$\,km\,s$^{-1}$ \citep{Coskunoglu2011}. However, since the proper motions are quite small (a few mas\,yr$^{-1}$), the correction for the solar motion has a large impact, especially on the P.A. of the motion. Interestingly, the P.A. of the corrected LSR motion is a worse match to the bow-shock orientation than the P.A. of the uncorrected motion (see Figs.~\ref{fig:cx_dra}, \ref{fig:pi_aqr}, and \ref{fig:phi_per}). Using $(U,V,W)_\sun = (10.00\pm0.36, 5.25\pm0.62, 7.17\pm0.38)$\,km\,s$^{-1}$ determined from the Hipparcos data by \citet{Dehnen1998} leads to $v_{\rm LSR}$ velocities which are slightly closer to the better matching heliocentric values. A similar discrepancy between bow-shock orientation and LSR motion (and less so with heliocentric motion) is found by \citet{Peri2012,Peri2015} for a large number of O- and B-type stars with bow shocks, as collected in the WISE E-BOSS survey. For this reason, we overplotted both the heliocentric and LSR motions on the WISE images in Figs.~\ref{fig:cx_dra}, \ref{fig:pi_aqr}, and \ref{fig:phi_per}. 
\begin{figure}[t!]
  \centering
  \includegraphics[width=9cm]{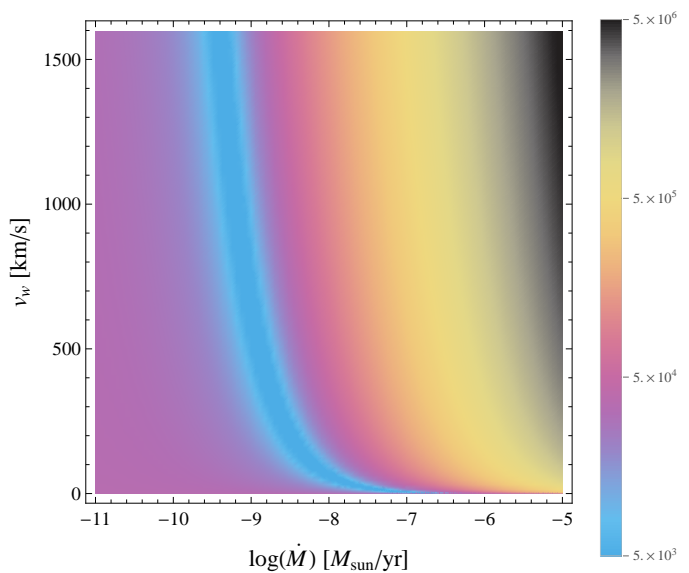}
  \caption{Density plot relating the mass-loss rate and wind velocity of $\pi$~Aqr to the bow-shock stand-off distance $R_0$. The colours indicate the difference in au to the measured $R_0= 36\,000$\,au.}
  \label{fig:mlr}
\end{figure}

\citet{Sarna1993}, \citet{vanRensbergen2011}, and more recently \citet{Deschamps2013,Deschamps2015} have suggested that Algols lose  a significant fraction of their initial mass during the mass-transfer phase. \citet{vanRensbergen2011} state that a hotspot mechanism may be invoked to remove up to $15\%$ of the binary-system initial mass. In this scenario, those parts of the stellar surface hit by the RLOF stream (the hotspot) emit radiation whose pressure triggers mass loss at a rate of up to $10^{-5}\,M_\sun$\,yr$^{-1}$.  Such a high mass-loss rate would  change the size of the bow shock considerably. Moreover, a stream winding around the binary system forms (see Fig.~1 of D15), possibly altering  the shape of the bow shock as well.

In order to identify the origin of the mass causing the observed bow shocks, one has to evaluate the dynamical age of the bow shock. The distance of the bow shocks to the central system is 22\,000\,au and 36\,000\,au for $\varphi$~Per and $\pi$~Aqr, respectively. If the bow shocks are caused by systemic mass loss triggered by a hot spot, the wind velocity is about 1000\,km\,s$^{-1}$ (see Fig.~3 of D15) and the resulting kinematic ages of the bow shocks are of the order of 100\,yr, much shorter than the duration of the mass-transfer phase in Algol systems \citep[$10^5$\,yr;][]{Deschamps2013}. Therefore, for Algols currently in the rapid mass-transfer phase, if a bow shock is present it will stay for the whole duration of the mass transfer.

Since the wind velocities of $\pi$~Aqr and $\varphi$~Per are not known, we cannot directly use Eq.~\ref{R0_eq} to relate the bow-shock stand-off distance to the mass-loss rate causing the bow shock and then compare the latter with the predictions of systemic mass-loss rates from the hotspot scenario of D15. Nevertheless, the relation between the mass-loss rate and the wind velocity may help us to evaluate the likelihood of systemic mass loss triggered by the hotspot scenario. In these calculations, we used the ISM densities $\rho_0 = n_{\rm H} \times m_{\rm H}$ and space velocities $v_{\rm *,LSR}$ listed in Table~\ref{tab:pa} with stand-off distances of $R_0=36\,000$\,au for $\pi$~Aqr and $R_0=22\,000$\,au for $\varphi$~Per, as measured on the WISE images.  Fig.~\ref{fig:mlr} depicts the relationship between mass-loss rate and wind velocity for $\pi$~Aqr. The colours show the differences in the calculated $R_0$ to the observed value of 36\,000\,au. 

For an expected wind velocity between 700 and 1500\,km\,s$^{-1}$, only a small range of mass-loss rates from $4\times10^{-10}\,M_\sun$\,yr$^{-1}$ to $4\times10^{-9}\,M_\sun$\,yr$^{-1}$ can match the observed stand-off distance ($\pm5000$\,au) of the bow shock around $\pi$~Aqr. This range of mass-loss rates is well below the systemic mass-loss rate inferred from the hotspot scenario (of the order of $10^{-5}\,M_\sun$\,yr$^{-1}$). Such a high mass-loss rate in combination with a wind velocity of $\approx1000$\,km\,s$^{-1}$ would cause a stand-off distance that is a factor of 100 larger than observed. To bring the wind velocity in line with the proposed $\dot{M}$ from the hotspot scenario and with the observed $R_0$, it has to decrease almost to 0. In other words, it is highly unlikely that a systemic mass loss of the order of $10^{-5}\,M_\sun$\,yr$^{-1}$ is currently present in $\pi$~Aqr. If systemic mass loss is ongoing, it does not exceed $10^{-8}\,M_\sun$\,yr$^{-1}$. We emphasise, however, that the mass-loss rate of $\pi$~Aqr inferred from the bow-shock properties fits well the value observed for the Be star \citep[$2.61\times10^{-9}\,M_\sun$\,yr$^{-1}$;][]{Snow1981}. We obtain similar results For $\varphi$~Per: $\dot{M}=2\times10^{-10}$ -- $6\times10^{-9}\,M_\sun$\,yr$^{-1}$ for $v_{\rm w}=$700--1500\,km\,s$^{-1}$ for a measured $R_0$ of $22\,000\pm5000$\,au.
\begin{figure}[t!]
  \centering
  \includegraphics[width=9cm]{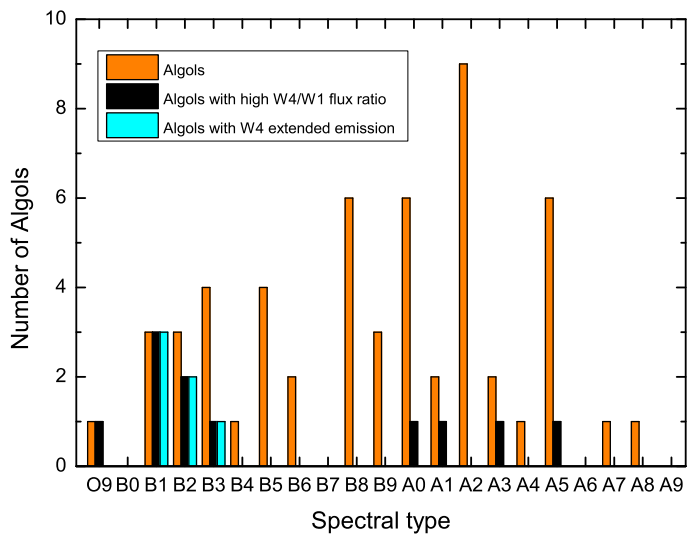}
  \caption{Distribution of Algols among early spectral types (orange bars), along with those with detected extended emission in the WISE W4 band (cyan bars) and with W4/W1 flux ratios falling above the black-body line (black bars).}
  \label{Fig:dist_spec1}
\end{figure}

\begin{figure}[t!]
  \centering
  \includegraphics[width=9cm]{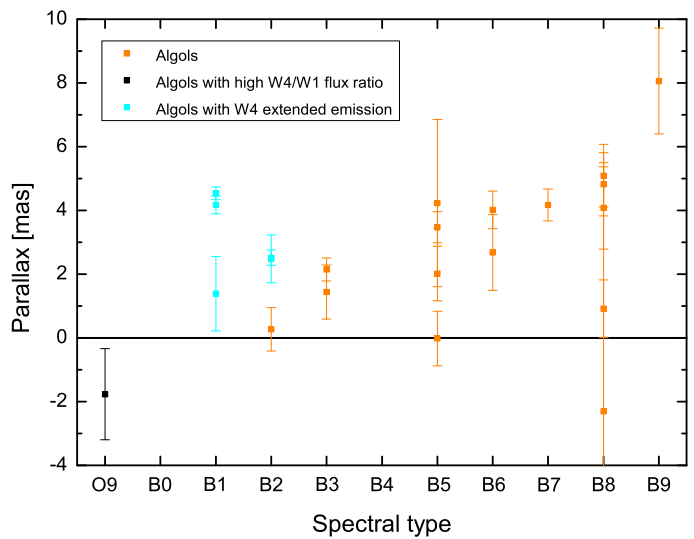}
  \caption{Distance distribution of Algols earlier than spectral type A. Orange squares are Algols without detection of circumstellar material, cyan squares show those with detected extended emission in the WISE W4 band, and black squares those with W4/W1 flux ratios falling above the black-body line (according to Fig.~13 of D15).}
  \label{Fig:dist_spec2}
\end{figure}

The Be star wind as the origin of the bow shock is further demonstrated by the fact that the bow shocks discovered in the present study are restricted to early-type B stars (Figs.~\ref{Fig:dist_spec1}--\ref{Fig:dist_spec2}). Our sample also includes  a O9.7Ibe star, \object{RY~Sct}, which shows a high WISE W4/W1 flux ratio. However, RY~Sct is much farther away than the three B stars discussed in the present paper since it has a parallax not significantly different from zero \citep{vanLeeuwen2007}; hence, if present, the extended emission would be hardly detectable. Given that the $V$ magnitude of RY~Sct (O9.7Ibe) amounts to 9.1,  compared to $V = 4.6$ for $\pi$~Aqr (B1Ve), RY~Sct is located at least a factor of 10 farther  away than $\pi$~Aqr. All other parameters being equal, the bow shock of RY~Sct would thus only extend up to $15\arcsec$  from the star, and would not be resolved by WISE. This conclusion is supported by Fig.~\ref{Fig:dist_spec2}, which reveals that the Be stars with a detected bow shock are the nearest and the warmest in the sample.

\begin{figure}[t!]
  \centering
  \includegraphics[width=9cm]{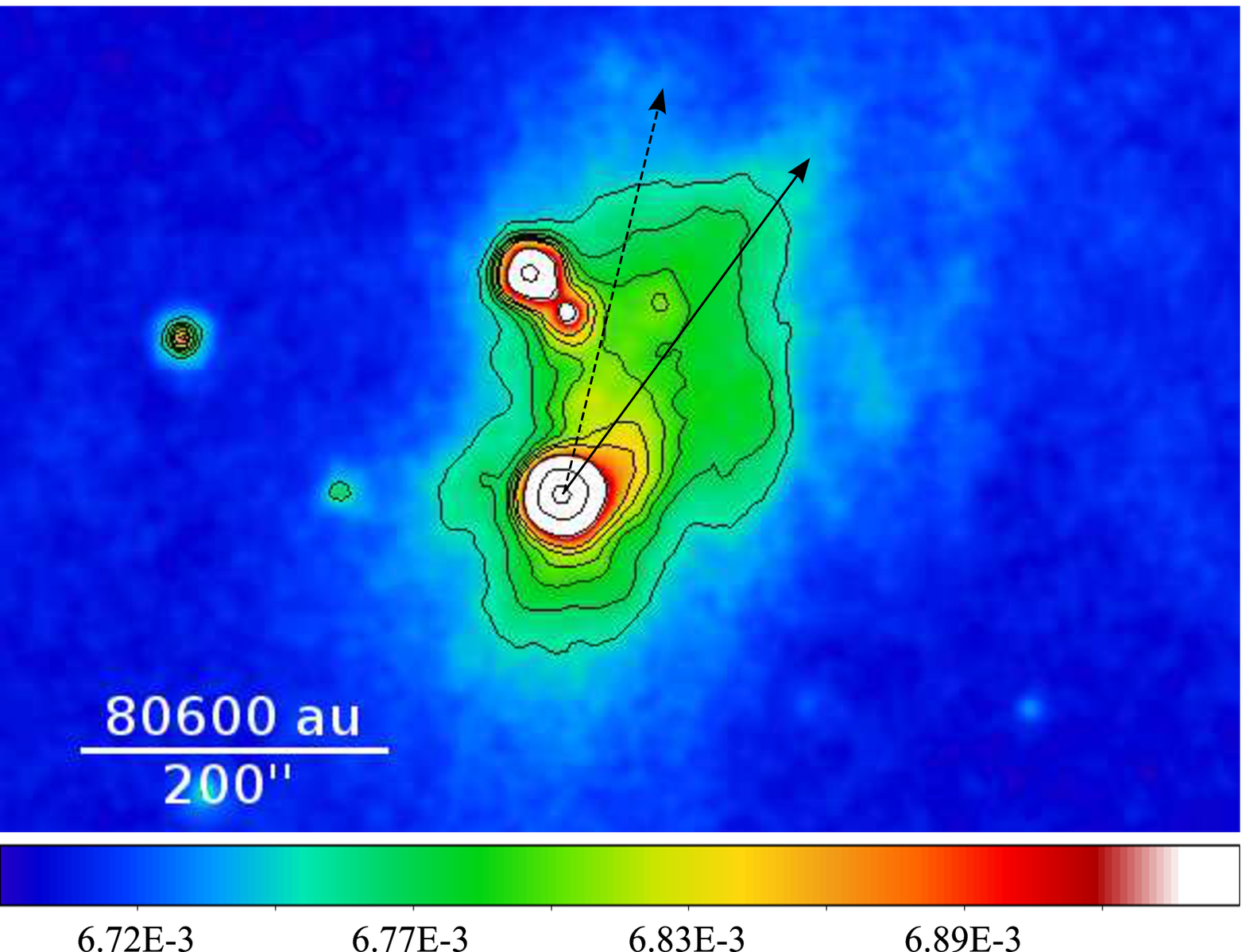}
  \includegraphics[width=9cm]{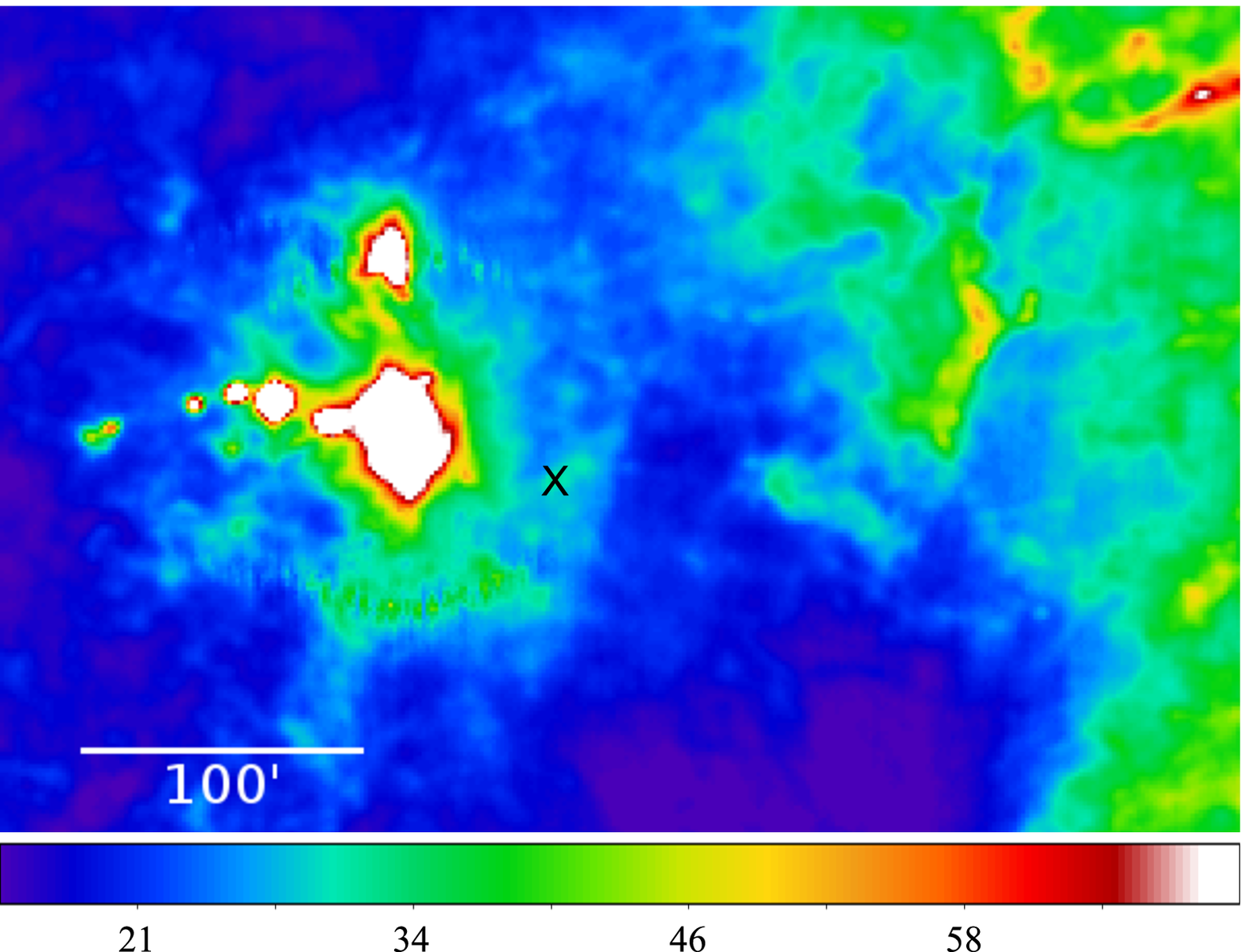}
  \caption{\textit{Upper panel:} WISE W4 image of V696~Mon at $22\,\mu$m. The continuous black arrow gives the uncorrected proper motion from the reprocessed Hipparcos catalogue \citep{vanLeeuwen2007}, while the dashed arrow points to the direction of the space motion corrected from the solar motion \citep{Coskunoglu2011}. The colour bar values are given in Jy\,pix$^{-1}$. \textit{Lower panel:} IRAS $100\,\mu$m image of the surroundings of V696~Mon (marked as an X). The colour bar values are given in MJy\,sr$^{-1}$.}
  \label{fig:V696_Mon}
\end{figure}

The second Be star with extended emission that we did not include in this study is the B2Vne star \object{V696~Mon}. The WISE W4 image is depicted in the upper panel of Fig.~\ref{fig:V696_Mon} and shows a peculiar morphology around the star. The extended emission reaches north of V696~Mon and seems to engulf the star \object{BD\,-06$\degr$1393} located $148\arcsec$  from V696~Mon at PA~$=8\degr$. This association is probably not real, since the Tycho-1 parallaxes of the two stars are quite different. Although the space velocity of V696~Mon [$v_{\rm *,LSR} = (15.8\pm6.2)$\,km\,s$^{-1}$] is comparable to the three stars studied here and the IR emission seems to be aligned with the direction of the space motion, the upstream structure does not resemble a bow shock. 

We also note that the surrounding ISM of V696~Mon is extremely rich (the star-forming region  Monoceros R2 is  $56\arcmin$ away) as seen in the IRAS 100\,$\mu$m image (lower panel of Fig.~\ref{fig:V696_Mon}), and it is difficult to conclude whether the WISE 22\,$\mu$m emission originates from CSM or ISM.

\section{Conclusion}
\label{sec:con}

We studied a sample of 70 Algol and Algol-like Be systems with entries in the WISE catalogue\footnote{We checked  the Herschel Science Archive as well, but only 2 of the 70 targets had been observed by the Herschel Space Observatory, neither of which showed circumstellar emission.} with the aim of identifying those Algol systems surrounded by dust left over by systemic mass loss. In D15, the two objects, CZ~Vel and SX~Aur, were discussed, and here we find that three new objects, CX~Dra, $\pi$~Aqr, and $\varphi$~Per,  show circumstellar material in the WISE W4 band at $22$\,$\mu$m;  $\pi$~Aqr and CX~Dra also show material in WISE W3 at 12\,$\mu$m. The two objects $\pi$~Aqr and  $\varphi$~Per show clear evidence of an interaction of circumstellar material with the ISM. For these events, we used the distance of the star to the bow shock to derive the mass-loss rate of the matter that escapes from the system. For both systems we estimate a mass-loss rate of $\approx 10^{-9}\,M_\sun$\,yr$^{-1}$, which is comparable to the mass-loss rate found for single B-type stars and much lower than the predicted systemic mass-loss rate from the hotspot scenario \citep[$\sim 10^{-5}\,M_\sun$\,yr$^{-1}$,][]{vanRensbergen2011,Deschamps2013,Deschamps2015}. Incidentally, several bow shocks have been reported around single Be stars by e.g. \citet{Kobulnicky2012}, \citet{Peri2012,Peri2015}, and \citet{Noriega1997}. It seems  most likely, therefore, that the bow shocks reported here are unrelated to the binary nature of the considered Be stars, and are of the same origin as those observed among non-binary Be stars. However, this does not entirely exclude the existence of systemic mass loss in these systems. It could well be that the mass transfer episode has already come to an end and the systemic mass-loss rate has dropped. 

Therefore, the prospects of finding observational support for systemic mass loss in Algols seem dark. Of the initial sample of 70 targets, only a handful of stars appear to exhibit IR excesses (see Figs.~\ref{fig:cx_dra}--\ref{fig:phi_per} and Figs.~13-14 of D15). It is conceivable that the simulations of D15 overestimate the amount of dust that survives the hotspot mechanism. For the two stars that show the presence of circumstellar material expressed by a bow shock ($\pi$~Aqr and $\varphi$~Per), neither shock  appears to be tied to systemic mass loss. 

In the case of CX~Dra, we were not able to identify the shaping mechanism responsible for the asymmetric circumstellar emission. Both common triggers for asymmetries, ISM interaction forming a bow shock and binary interaction forming an Archimedean spiral, can be excluded for various reasons. We note therefore that this system is an interesting case for further observations since it may be a case where mass lost from the system is visible in the circumstellar environment.

\begin{acknowledgements}
We want to thank the anonymous referee for the constructive and helpful remarks. This research is supported by the Belgian Federal Science Policy office via the PRODEX Program of ESA. AM acknowledges funding by the Austrian Science Fund FWF under project numbers P23586 and P23006-N16 and by the Austrian Research Promotion Agency FFG under project number FA 538019. RD acknowledges support from the Communaut\'{e} fran\c{c}aise de Belgique -- Actions de Recherche Concert\'{e}es and benefits from a European Southern Observatory studentship. We made use of the NASA/IPAC Infrared Science Archive, which is operated by the Jet Propulsion Laboratory, California Institute of Technology, under contract with the National Aeronautics and Space Administration and of the SIMBAD database, operated at CDS, Strasbourg, France.
\end{acknowledgements}

\bibliographystyle{aa}
\bibliography{paper}

\begin{appendix}
\section{List of objects}
\label{objectlist}

\vspace{0.5cm}
\tiny
\topcaption{Seventy Algols and Algol-like Be stars with a WISE source counterpart sorted by declination. Objects marked with an asterisk (*) show extended emission in WISE W4, while the plus sign (+)  marks objects with a WISE W4/W1 flux ratio which is above the black-body law (see Tab.~6 in D15). References: (1): \citet{Harmanec2001}; (2): \citet{Brancewicz1980}; (3): \citet{Budding2004}.} 

\tablefirsthead{\toprule\toprule  V* Name &  2MASS  & HD/BD  & Spec. type & Ref. \\ \midrule}
\tablehead{%
\multicolumn{5}{c}%
{{\bfseries  Continued from previous column}} \\
\toprule
 V* Name &  2MASS & HD/BD & Spec. type & Ref. \\ \midrule}
\tabletail{%
\midrule \multicolumn{5}{r}{{Continued in next column}} \\ \midrule}
\tablelasttail{%
\\ \bottomrule}

\begin{supertabular}{lrrrr}
BP Mus                          &        J12503772-7146186      & & & 3          \\
DW Aps                          &        J17233003-6755448      &       HD 156545 & B6III & 2,3    \\ 
EP TrA                          &        J15492615-6415574      &       HD 140809 & A0 & 2,3       \\
AN Tuc                          &       J23302225-5825346       &       HD 221184 & A5III & 2,3    \\R Ara                                         &       &       HD 149730 &  B9IV & 1,2,3  \\
UZ Nor +                        &       J16281156-5319215       & & B?  & 3       \\
V646 Cen                        &       J11365877-5312354       &       HD 100987 & B8IV & 2,3     \\RV Pic                        &       J04572970-5208458       &       HD 32011 & A1V & 2,3       \\
CZ Vel *+                               &       J09104446-5042405       & & B3 &  3        \\
KV Pup                          &       J07471912-4832122       &       HD 63562 & A0IV & 2,3      \\
TT Hor                          &        J03270438-4552566      &  & &  3        \\
DN Vel                          &        J09193768-4540477      &       HD 80692 & A0III & 2,3     \\
VY Mic                          &        J20490707-3343543      &       HD 198103 & A4III & 2,3     \\
DM Pup                          &        J08070409-2531522      &  & A2.5 & 3             \\
AA Pup                          &        J08013612-2443034      &       HD 66226 & F3IV & 2,3      \\YY CMa                        &        J07005186-1914315      &  & A2V & 3              \\
AO Eri                          &        J04320093-1744475      &       & A2 & 2,3        \\
SS Lep  +                                       &        J06045913-1629039      &       HD 41511 & A1V & 1 \\
W Ser +                         &        J18095070-1533009      &       HD 166126 & F5III & 1,2,3  \\
RY Sct +                                                &        J18253147-1241241      &       HD 169515 & O9.7Ibep & 1,2,3       \\
XY Pup                          &       &               HD 67862 & A3 & 1,2,3         \\
V644 Mon                        &        J06570938-1049281      &       HD 51480 & Ape & 1         \\
AW Mon                          &       &               BD-10 2233 & A2 & 2,3     \\
RZ Sct                          &        J18263352-0912060      &       HD 169753 & B3Ib & 1,2,3   \\
XZ Aql                          &        J20221335-0721034      &       HD 193740 & A2 & 2,3        \\
V696 Mon *+                                     &        J06041349-0642321 &       HD 41335 & B2Vne & 1  \\
AR Mon                          &        J07204845-0515357      &       HD 57364 & K0II & 2,3      \\AU Mon                        &        J06545471-0122328      &       HD 50846 & B4IV & 1,2,3    \\
V509 Mon +                      &        J06471071-0102147      &       & G4IV & 3        \\
$\pi$ Aqr *+                                    &        J22251662+0122389      &       HD 212571 & B1Ve & 1  \\
AC Tau                          &        J04370635+0141311      &       & A8 & 2,3        \\
SS Cet                          &               &       HD 17513 & A2 & 2,3  \\
AX Mon                          &        J06303293+0552012      &       HD 45910   & B2III & 1  \\
DN Ori                          &        J06002835+1013049      &       HD 40632   &  A2e &  1,2,3  \\
FM Ori                          &        J05085439+1033341      &       HD 241071 & A5 & 3         \\
V930 Oph                        &        J18414565+1202111      & & & 3 \\
BI Del                          &        J20273862+1420091      &       & G0 & 2,3        \\
AL Leo                          &        J09581290+1817282      &       & F5 & 2,3        \\
U Sge                           &        J19184840+1936377      &       HD 181182 & B7.5V & 1,2,3  \\
AL Gem                          &        J06573855+2053325      &       HD 266913 & F6V & 2,3      \\RS Vul                                                &        J19174000+2226284       &       HD 180939 & B5V & 2,3   \\
DH Her +                        &        J18473455+2250458      &       HD 343047 & A5 & 2,3        \\RW Tau                       &        J04035432+2807334      &       HD 25487    & B8Ve & 1,2,3  \\
U CrB                           &        J15181133+3138492      &       HD 136175 & B6V & 1,2,3    \\
BC Aur                          &        J05461654+3250500      &       & & 3      \\
RX Gem                          &        J06501154+3314207      &       HD 49521    & A0 & 1,2,3  \\
CG Cyg                          &        J20581343+3510298      &       & G9.5V & 2,3     \\
V367 Cyg +                      &        J20475958+3917156      &       HD 198288 & A3Ibep & 1,2   \\
AB Per                                          &        J03374520+4045494      &       HD 275604  & A5 & 2,3  \\
SX Aur *+                       &        J05114292+4209553      &       HD 33357    & B1Vne & 1,2,3  \\
RW Per                          &        J04201676+4218517      &       HD 276247  & A2 & 1,2,3  \\
TX UMa                          &        J10452050+4533586      &       HD 93033    & B8V & 1,2,3  \\
GK And                          &        J23534719+4534458      &       & & 3     \\
V995 Cyg                        &        J19483443+4613424      & & B8 &3         \\
SW Cyg                          &        J20065793+4617581      &       HD 191240 & A2 & 1,2,3 \\IM Aur                    &        J05152973+4624214      &       HD 33853 & B9 & 2,3        \\
RY Per                          &        J02454210+4808379      &       HD 17034    & B8V & 1,2,3  \\
KX And                          &        J23070621+5011324      &       HD 218393 & B3pe   & 1  \\$\varphi$ Per *+                         &        J01433964+5041192       &       HD 10516 & B1.5Ve & 1   \\
AY Per                          &        J03102513+5055543      &       HD 232756 &        B9 & 2,3  \\
CX Dra *+                                               &        J18464309+5259166      &       HD 174237 & B2.5Ve & 1,3   \\
V442 Cas +                      &        J23401479+5357339      &       & A0 & 3  \\
SX Cas                          &        J00104207+5453293      &       HD 232121 & B5 & 1,2,3     \\
DM Per                          &        J02255800+5606099      &       HD 14871 & B5V & 2,3       \\
GG Cas                          &       &       BD+55 274       & B5 & 2,3  \\
RX Cas                          &        J03074573+6734387      &       & A5III & 2,3  \\
XY Cep                          &        J23523291+6856015      &       & B8 & 2,3        \\
SS Cam                          &        J07162474+7319570      &       & G1III & 2,3     \\
XZ Cam                          &        J05171266+7550053      &       & A0 & 2,3        \\
RS Cep                          &        J05060320+8014524      & & A5III & 1,2,3         \\
TY UMi                          &        J15175751+8351340      &       HD 138818 & F0 & 3 
\end{supertabular}%

\end{appendix}
\end{document}